\documentclass[12pt]{article}
\usepackage{amsmath}
\usepackage{comment}
\usepackage{authblk}
\usepackage[a4paper, margin=1in]{geometry}
\usepackage{graphicx} 
\usepackage{ragged2e}
\usepackage{amsmath}
\usepackage{amssymb}
\usepackage{braket}
\usepackage{algorithm}
\usepackage{algpseudocode}
\usepackage{commath}
\usepackage{siunitx}
\usepackage{tabularx}
\usepackage{array}
\usepackage{longtable}
\usepackage{float}
\restylefloat{table}
\usepackage{multirow}
\usepackage{soul}
\usepackage[normalem]{ulem}
\usepackage[backend=biber,style=numeric,sorting=none,doi=false,isbn=false,url=false]{biblatex}
\usepackage{mathtools} 
\usepackage[colorlinks=true, allcolors=black]{hyperref}
\newcolumntype{Y}{>{\centering\arraybackslash}X}
\newbibmacro{string+doi}[1]{%
  \iffieldundef{doi}{\iffieldundef{url}{#1}{\href{\thefield{url}}{#1}}}{\href{http://dx.doi.org/\thefield{doi}}{#1}}}	
\DeclareFieldFormat*{title}{\usebibmacro{string+doi}{\mkbibemph{\textcolor{blue}{#1}}}}

\providecommand{\keywords}[1]
{
  \small	
  \textbf{\textit{Keywords---}} #1
}

\title{Secure and Efficient n-Qubit Entangled State Teleportation Using Partially Entangled GHZ Channels and Optimal POVM}
\author[1,$\dagger$]{Animesh Banik}
\author[1,$\dagger$]{Md. Shihab Khan}
\author[2,$\dagger$]{Rafid Masrur Khan}
\author[3]{Syed Emad Uddin Shubha}
\author[2,4,*]{Mahdy Rahman Chowdhury}
\affil[1]{Department of Physics, University of Chittagong, Chittagong, Bangladesh}
\affil[2]{Department of Electrical and Computer Engineering, North South University, Dhaka, Bangladesh}
\affil[3]{Division of Computer Science, Louisiana State University, USA}
\affil[4]{NSU Center of Quantum Computing, North South University}
\affil[$\dagger$]{Mahdy Research Academy, Dhaka, Bangladesh}
\affil[*]{Corresponding author}
\date{}

\numberwithin{equation}{section}
\usepackage{float}
\begin{document}
\maketitle
\begin{center}
\small
\textit{Author accepted manuscript. Published in AVS Quantum Science 7, 035802 (2025). \\
DOI: \href{https://doi.org/10.1116/5.0284072}{10.1116/5.0284072} \\
© 2025 AIP Publishing. This version is made available under AIP Publishing's web posting policy.}
\end{center}
    
\begin{abstract}
\sloppy
    We introduce an efficient and versatile quantum teleportation protocol for specific types of $n$-qubit entangled states. By employing a partially entangled Greenberger-Horne-Zeilinger (GHZ) state as the quantum channel and an optimal Positive Operator-Valued Measure (POVM) based on an improved reciprocal state formulation, we achieve unambiguous state discrimination. The scheme has been generalized to support various entangled state configurations and demonstrates a notable reduction in classical communication costs for these states compared to standard Bell-basis teleportation. Its capacity for integration with conventional protocols is pivotal, enhancing quantum hop-by-hop communication security by allowing strategic choices in quantum channel and teleportation strategy.
    
\end{abstract}
\keywords{ Quantum teleportation, entangled state, GHZ, POVM, unambiguous state \mbox{discrimination}, quantum hop-by-hop communication}

\section{Introduction}
Quantum teleportation is a fundamental component of quantum communication systems, enabling the transfer of quantum information between spatially separated parties without physically transmitting the particles themselves \cite{Bennett1993}. First proposed by Bennett et al. in 1993, the experimental feasibility of this concept was explored shortly thereafter \cite{bennett1995towards}, leading to its first experimental realization by Bouwmeester et al. in 1997 using photon polarization states \cite{bouwmeester1997experimental}. Since these pioneering works, teleportation has been demonstrated across a variety of physical systems, including trapped ions \cite{riebe2004deterministic} and atomic ensembles \cite{ursin2007entanglement}, and over increasingly significant distances via free-space optical channels (up to 100 km \cite{Yin2012QuantumTA}) and satellite-based platforms \cite{ren2017ground}. These advancements have established teleportation as a cornerstone for emerging technologies such as the quantum internet \cite{valivarthi2020teleportation}, integrated space-ground quantum communication networks \cite{chen2021integrated}, and communication between non-neighbouring nodes in quantum networks \cite{hermans2022qubit}.

\par Despite these advancements, existing teleportation schemes face challenges when scaled to multipartite entangled states, particularly with respect to resource efficiency and classical communication overhead. The Greenberger–Horne–Zeilinger (GHZ) state, introduced by Greenberger et al. in 1989 \cite{greenberger1989going} and formally characterized in 1990 \cite{greenberger1990bell}, generalizes Bell-type correlations to multipartite systems and is fundamental in many multi-qubit quantum information protocols. While offering richer entanglement structures, protocols based on GHZ states often require more sophisticated strategies for state discrimination and reconstruction compared to those using bipartite entanglement. Quantum state discrimination thus plays a critical role, determining how measurement outcomes map to state recovery. Foundational works by Helstrom \cite{helstrom1969quantum}, Ivanovic \cite{ivanovic1987differentiate}, Dieks \cite{dieks1988overlap}, and Peres \cite{peres1988differentiate} laid the theoretical basis for distinguishing quantum states, with Chefles \cite{chefles1998unambiguous} later establishing that linearly independent states can be perfectly distinguished using generalized measurements.The work of Mor and Horodecki \cite{mor1999teleportation} extended teleportation to generalized measurements (POVMs) and introduced the notion of conclusive (probabilistic) teleportation. In a foundational approach, the Horodecki family \cite{horodecki1999general} introduced the singlet fraction as the key figure of merit for quantum teleportation and analyzed general teleportation channels. Zhang et al. \cite{zhang2022revisiting} introduced a new approach to state discrimination, claiming more information recovery than the standard unambiguous state discrimination. Contemporarily, the development of systematic methods for constructing POVMs that optimally discriminate between non-orthogonal quantum states under different error and inconclusive outcome trade-offs was undertaken by ding et al.\cite{ding2021povm}. Recently, in \cite{gupta2024unambiguous} it has been shown that optimal unambiguous sequence discrimination can be achieved through individual measurements, eliminating the need for collective strategies, When the parent set elements share identical real inner products.

\par These theoretical advancements in state discrimination led to the adoption of Positive Operator-Valued Measures (POVMs) in quantum protocols, particularly for scenarios involving non-orthogonal states that arise with partially entangled channels. POVMs offer a more general measurement framework than standard projective measurements, enabling more efficient state identification in certain contexts. Eldar et al.\cite{eldar2003designing} showed that for a pure-state ensemble represented by rank-one density operators (possibly dependent), the optimal measurement is likewise a pure-state measurement composed of rank-one operators. Consequently, Eldar \cite{eldar2003semidefinite} provided a significant contribution by formulating a semidefinite programming approach to design optimal POVMs for unambiguous state discrimination.  The idea of probabilistic teleportation was first introduced by \cite{li2000probabilistic} and then extended through the works \cite{son2001conclusive,agrawal2002probabilistic} of Son et al.,Agarwal and Pati. Building on such concepts, previous efforts like Wang et al. \cite{wang2018two} demonstrated a probabilistic teleportation protocol for two-qubit entangled states using a partially entangled GHZ state and POVMs optimized for unambiguous discrimination, marking a key step in integrating complex entanglement structures with advanced measurement strategies. In a complementary line of research, Zhang \cite{ZHANG2006qsscm} explored the teleportation of n-qubit information using quantum secret sharing, though this primarily focused on security aspects rather than resource efficiency for complex multipartite quantum states in the same vein.

\par In this paper, we employ an n-qubit generalization of the scheme by Wang et al. \cite{wang2018two} to sustain the teleportation of an n physical qubit quantum state that is the encoded form of a logical qubit. The specific nature of our chosen input state and an n+1 qubit partially entangled GHZ channel results in a significant improvement on the efficiency in terms of communication overhead. However, the intriguing aspect of the proposed scheme lies in the resilience of the input state during teleportation as compared to the popular protocol proposed by Bennett et al. \cite{Bennett1993}. We boast of not the generality of the initial input state to be teleported but of the applicability of this protocol in the physical systems where X-type i.e. bit-flip type errors are main contributors to noise. This focused approach, serving as a basic step, paves the way for a complementary protocol for phase errors which upon combining with the newly proposed scheme would be analogous to the Shor code. This potential protocol capable of protecting against arbitrary single-qubit errors is of great importance considering the future experimental realizations of quantum teleportation of protected states in distributed quantum network.

\section{Background and Technical Overview}
The preceding introduction has outlined the significance of quantum teleportation, the challenges in scaling protocols for multipartite states, and the context for the contributions of this paper. To further elaborate on the underpinnings of our proposed n-qubit teleportation scheme, this section now delves into a more focused discussion of the key physical principles and established methodologies that are integral to its design and operation. This includes a detailed examination of quantum entanglement, exemplified by both Einstein-Podolsky-Rosen (EPR) pairs and Greenberger-Horne-Zeilinger (GHZ) states; an overview of the mechanics of the standard quantum teleportation protocol; and an exploration of quantum measurement theories, contrasting projective measurements (PVM) with the more generalized Positive Operator-Valued Measures (POVM). A thorough understanding of these foundational elements is essential for appreciating the advancements and nuances of the protocol presented subsequently.

\subsection{Quantum Entanglement: EPR Pairs and GHZ States}

\textbf{Entanglement} is a non-local property of quantum mechanics where two or more quantum systems are interconnected in such a way that their individual quantum states cannot be described independently, regardless of the distance separating them. This phenomenon is a primary resource in quantum communication.

The simplest and most well-known form of bipartite entanglement is the \textbf{Einstein-Podolsky-Rosen (EPR) pair}, often referred to as a Bell state. For example, one of the four maximally entangled Bell states for two qubits can be written as:
$$|\Phi^+\rangle_{12} = \frac{1}{\sqrt{2}}(|0_10_2\rangle + |1_11_2\rangle)$$
Here, qubits 1 and 2 are perfectly correlated. Measuring qubit 1 in the computational basis and obtaining $|0\rangle_1$ instantly means qubit 2 is in state $|0\rangle_2$, and vice-versa, even if 1 and 2 are light-years apart.

Beyond two-qubit systems, \textbf{multipartite entanglement} involves three or more qubits. A prominent example is the \textbf{Greenberger-Horne-Zeilinger (GHZ) state}. For n qubits, a maximally entangled GHZ state can be expressed as:
$$|GHZ\rangle_n = \frac{1}{\sqrt{2}}(|0\rangle^{\otimes n} + |1\rangle^{\otimes n})$$
This state exhibits strong n-partite correlations. The GHZ state was first conceptualized by Greenberger, Horne, and Zeilinger. The teleportation protocol explored in this paper utilizes a specific form of an $(n+1)$-qubit GHZ state as a quantum channel, which is generally a pure \textbf{partially entangled GHZ state}:
$$|GHZ\rangle_{\text{channel}} = a|0\rangle^{\otimes n+1} + b|1\rangle^{\otimes n+1}$$
where $|a|^2 + |b|^2 = 1$, and the degree of entanglement depends on the coefficients $a$ and $b$.

\subsection{The Standard Quantum Teleportation Protocol}

The original quantum teleportation protocol, introduced by Bennett et al. \cite{Bennett1993} in 1993, describes how to transmit an unknown arbitrary single-qubit quantum state from a sender (Alice) to a receiver (Bob).

Let the unknown quantum state Alice wishes to teleport be $|\psi\rangle_3 = \alpha|0\rangle_3 + \beta|1\rangle_3$. The protocol proceeds as follows:
\begin{enumerate}
    \item \textbf{Shared Entanglement:} Alice and Bob must pre-share an EPR pair, for instance, the $|\Phi^+\rangle_{12}$ state mentioned above. Alice possesses qubit 1 and Bob possesses qubit 2.
    \item \textbf{Alice's Joint Measurement:} Alice performs a joint measurement on her qubit 3 (carrying the unknown state $|\psi\rangle_3$) and her qubit 1 from the EPR pair. This measurement is typically done in the Bell basis, which projects the two qubits onto one of the four Bell states.
    \item \textbf{Classical Communication:} Alice sends the outcome of her Bell measurement (which can be one of four possibilities, requiring 2 classical bits of information) to Bob over a classical communication channel.
    \item \textbf{Bob's Unitary Operation:} Based on the 2 classical bits received from Alice, Bob applies one of four specific unitary operations: \( I \), \( \sigma_x \), \( \sigma_z \), or $\sigma_x \sigma_z $ to his qubit \( B \).
 This transforms Bob's qubit into an exact replica of the original state $|\psi\rangle_3$.
\end{enumerate}
This process effectively transfers the quantum state from Alice to Bob without physically moving the qubit 3, consuming the initial entanglement of the EPR pair in the process.

\subsection{Quantum Measurement: PVM vs. POVM}

Quantum measurements are the means by which information is extracted from a quantum system. The nature of the measurement plays a crucial role in the outcomes and the post-measurement state.

\textbf{Projective Measurement (PVM):} Often called von Neumann measurements, these are described by a set of orthogonal projection operators $\{P_m\}$ acting on the Hilbert space of the system. These operators satisfy the completeness relation $\sum_m P_m = I$ (where $I$ is the identity operator) and orthogonality $P_m P_k = \delta_{mk} P_m$. If a system is in state $\rho$, the probability of obtaining outcome $m$ is given by $p(m) = \text{Tr}(\rho P_m)$. After the measurement, if outcome $m$ is obtained, the state of the system collapses to $\rho_m = \frac{P_m \rho P_m^\dagger}{\text{Tr}(\rho P_m)}$. Projective measurements are sufficient for distinguishing orthogonal quantum states perfectly.

\textbf{Positive Operator-Valued Measure (POVM):} POVM represents a more general framework for quantum measurements. A POVM is described by a set of positive semi-definite operators (POVM elements) $\{\Pi_m\}$ that sum to the identity operator, $\sum_m \Pi_m = I$. The probability of obtaining outcome $m$ when measuring a state $\rho$ is $p(m) = \text{Tr}(\rho \Pi_m)$. Unlike projective measurements, the POVM elements $\Pi_m$ are not required to be orthogonal projectors. This generalization allows POVMs to perform tasks impossible for PVMs, most notably in the context of distinguishing non-orthogonal quantum states.

When attempting to discriminate non-orthogonal states, POVMs can be designed for implementing \textbf{unambiguous state discrimination}. In this strategy, each outcome either perfectly identifies a state or is inconclusive, meaning no error is made in identification, though success is probabilistic. This is particularly relevant in scenarios where the quantum channel used for teleportation (like a partially entangled GHZ state) leads to Alice needing to distinguish between states that are not mutually orthogonal. The design of an optimal POVM aims to maximize the probability of successfully and unambiguously discriminating these states. Foundational work by Helstrom, Ivanovic, Chefles, Dieks, and Peres has contributed significantly to the understanding and application of state discrimination techniques, including those utilizing POVMs.

\section{The Proposed n-Qubit Teleportation Protocol}
The primary motivation for this work is to develop a robust method for transferring protected quantum states between nodes in a distributed quantum network. We specifically address the teleportation of the generalized $n$-qubit state, defined as:
$$ \ket{\chi_{st}} = \alpha \ket{t_1t_2...t_n} + (-1)^{s} \beta \ket{t_1^\prime t_2^\prime...t_n^\prime} $$
Here $s, t_i \in \{0,1\}$ and $t_i^\prime= 1-t_i$. The parameters $\alpha$ and $\beta$ represent the probability amplitudes. This state represents a single logical qubit encoded within a Pauli frame (defined by $s$ and $t_i$) using an $n$-qubit repetition code. Our protocol is tailored for physical systems where $X$-type (bit-flip) errors are the dominant noise source, as the repetition code basis is expressly designed for this contingency. This focus serves as a foundational step; by developing a parallel protocol for phase errors, the two methods could be concatenated, in a manner analogous to the Shor code, to protect against arbitrary single-qubit errors. The following protocol provides a direct and efficient method for teleporting these encoded logical states over realistic, partially entangled channels.

\subsection{Teleportation of n-qubit GHZ-like entangled state}

Alice and Bob are two nodes connected through a quantum channel. Alice intends to transmit an n-qubit entangled unknown state to Bob. We consider the form of the initial state as: 
\begin{equation}
    \ket{\chi_0} = \alpha\ket{0}^{\otimes n}+\beta\ket{1}^{\otimes n}
\end{equation}
The quantum channel used will be $(n+1)$ qubit pure partially entangled GHZ state \cite{greenberger1989going} of the form:
\begin{equation}
    \ket{GHZ} = a\ket{0}^{\otimes n+1}+b\ket{1}^{\otimes n+1}
\end{equation}
Here, $a,b$ represent the probability amplitudes of the quantum channel. By design, the following conditions must be satisfied for this scheme:
\begin{equation}
    \begin{aligned}
       \abs{a} \geq \abs{b}, \hspace{3cm} \abs{a}^2 + \abs{b}^2 = 1 
    \end{aligned}
\end{equation}
Note that the case of $a=b$ leads to a maximally entangled channel. The combined state can be written as,
\begin{equation}
    \ket{\psi_{sys}} = \ket{\chi_0}_{1,2,...,n}\otimes\ket{GHZ}_{n+1,n+2,...,2n+1} 
\end{equation}
Expanding the above equation, the system could be expressed as,
\begin{equation}\label{eq4}
    \begin{aligned}
        \ket{\psi_{sys}} = (\alpha\ket{0}^{\otimes n}+\beta\ket{1}^{\otimes n})_{1,2,...,n}
        \otimes(a\ket{0}^{\otimes n+1}+b\ket{1}^{\otimes n+1})_{n+1,n+2,...,2n+1}\\
        =\frac{1}{2}(a\ket{0}^{\otimes n}\otimes\ket{0}+b\ket{1}^{\otimes n}\otimes\ket{1})_{1,2,...,n+1}
        \otimes(\alpha\ket{0}^{\otimes n}+\beta\ket{1}^{\otimes n})_{n+2,n+3,...,2n+1}
        \\+\frac{1}{2}(a\ket{0}^{\otimes n}\otimes\ket{0}-b\ket{1}^{\otimes n}\otimes\ket{1})_{1,2,...,n+1}
        \otimes(\alpha\ket{0}^{\otimes n}-\beta\ket{1}^{\otimes n})_{n+2,n+3,...,2n+1}
        \\+\frac{1}{2}(a\ket{1}^{\otimes n}\otimes\ket{0}+b\ket{0}^{\otimes n}\otimes\ket{1})_{1,2,...,n+1}
        \otimes(\beta\ket{0}^{\otimes n}+\alpha\ket{1}^{\otimes n})_{n+2,n+3,...,2n+1}
        \\+\frac{1}{2}(a\ket{1}^{\otimes n}\otimes\ket{0}-b\ket{0}^{\otimes n}\otimes\ket{1})_{1,2,...,n+1}
        \otimes(\beta\ket{0}^{\otimes n}-\alpha\ket{1}^{\otimes n})_{n+2,n+3,...,2n+1}
    \end{aligned}
\end{equation}

\begin{figure}[h]
    \centering
    \includegraphics[width=01\linewidth]{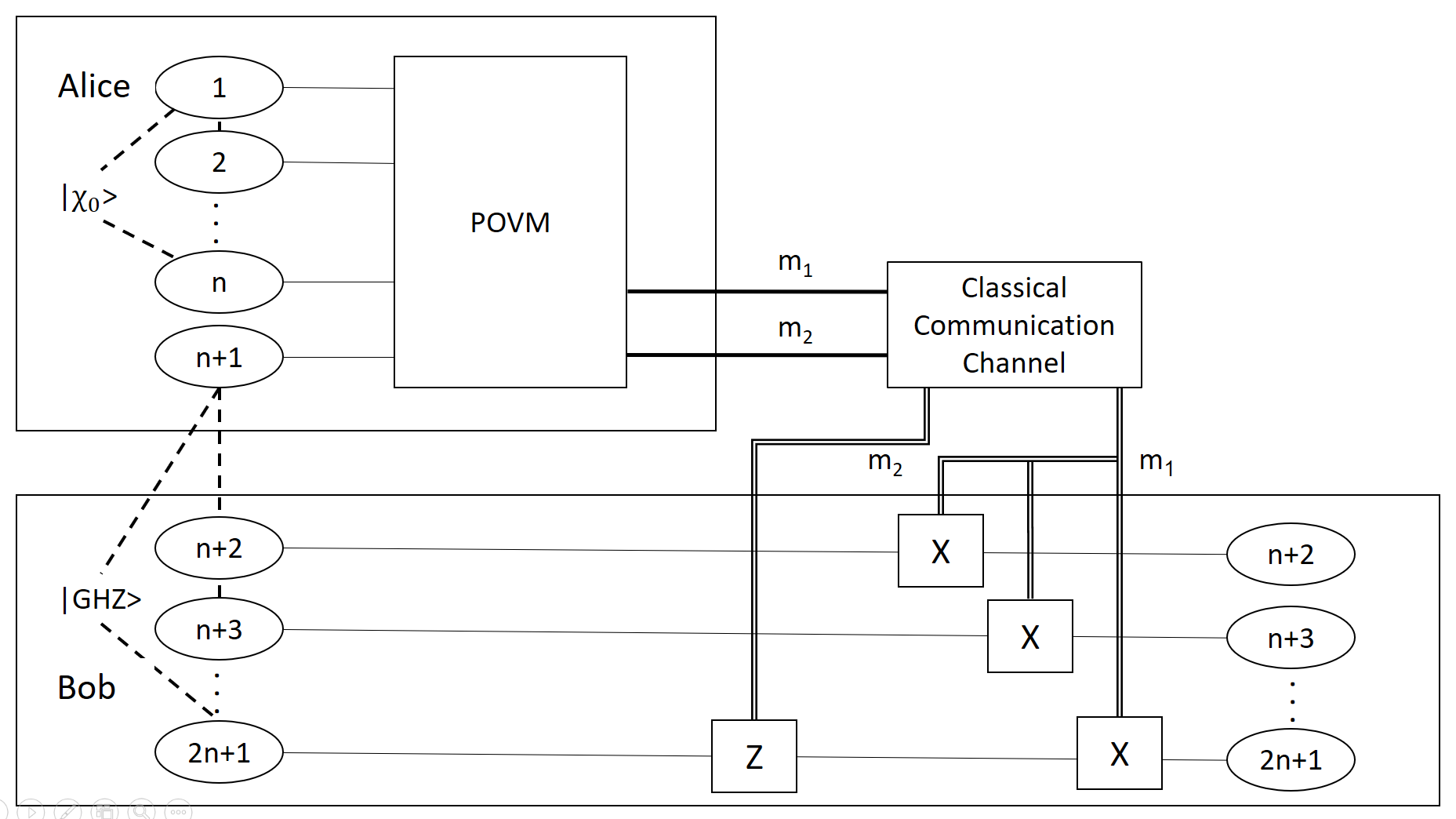}
    \caption{Schematic representation of the quantum teleportation protocol for an n-qubit GHZ-like entangled state. Alice holds the unknown n-qubit state $\ket{\chi_0}=\alpha \ket{0}^{\otimes n} + \beta \ket{1}^{\otimes n}$ and one qubit from the shared partially entangled GHZ state. She performs a Positive Operator-Valued Measurement (POVM) on her (n+1) qubits and transmits the classical outcomes  $m_1$ and $m_2$ to Bob via a classical communication channel. Bob, who holds the remaining n qubits, applies appropriate unitary operations (X and Z gates) based on the received classical bits as per Eq.~\ref{gen unitary} to reconstruct the original state $\ket{\chi_0}$.}
    \label{fig_chi_zero}
\end{figure}

For the scheme to succeed, Alice is required to do joint measurement on the particles 1,2,...,n+1 so that the state on particles n+2,n+3,...,2n+1 collapse into one of the four possible states. It is to be noted that the states of Alice's and Bob's particles have a correlation. Depending on the correct detection of state of Alice's qubits, Bob is required to do appropriate local operations on his qubits. This local operations then lead to the reconstruction of the original state. The whole protocol can be understood by following the Figure~\ref{fig_chi_zero}.\newline
Note that Alice must convey the measurement results to Bob through a classical channel before applying local operation on Bob’s end. The discrimination of the states in this case has been formulated as a typical unambiguous state discrimination problem \cite{ivanovic1987differentiate} using POVM \cite{dieks1988overlap, peres1988differentiate, chefles1998unambiguous} since the states in question are non-orthogonal in case of partially entangled quantum channel.

\subsubsection{Unambiguous state discrimination formulation}

From Eq.\ref{eq4}, the states to be discriminated can be written in the vector form as:
\begin{equation}
\begin{aligned}
    \ket{\phi_1}=[a,0,...,0,b]^T, \ket{\phi_2}=[a,0,...,0,-b]^T, \\
    \ket{\phi_3}=[0,b,...,a,0]^T, \ket{\phi_4}=[0,-b,...,a,0]^T
\end{aligned}
\end{equation}
Note that the dimension of the $\ket{\phi_i}$s is $2^{n+1}$. We find the following relation describing the inner products of the $\ket{\phi_i}$s:
\begin{equation}\label{phi_i_phi_j}
    \begin{aligned}
        \braket{\phi_i|\phi_j} = d_{ij} = \abs{a}^2 \begin{bmatrix}
            1& 1& 0& 0\\
            1& 1& 0& 0\\
            0& 0& 1& 1\\
            0& 0& 1& 1
        \end{bmatrix} + \abs{b}^2 \begin{bmatrix}
             1& -1&  0&  0\\
            -1&  1&  0&  0\\
             0&  0&  1& -1\\
             0&  0& -1&  1
        \end{bmatrix}
    \end{aligned}
\end{equation}
As per Eq.~\ref{phi_i_phi_j}, one can say that the $\ket{\phi_i}$s are orthogonal only if $a=b$. This only occurs when the quantum channel is maximally entangled. In case of partially entangled quantum channel, these four states are non-orthogonal. We design an optimal POVM in order to detect the state correctly. We accept the failure of this protocol when the measurement result is inconclusive. In that case Alice simply starts the protocol anew. \newline
Here, we use the reciprocal states of the $\ket{\phi_i}$s in order to design the POVM operators \cite{eldar2003semidefinite}. Finding the reciprocal state can be achieved using the following relation,
\begin{equation}\label{phi_i_tilde}
    \ket{\tilde{\phi_i}} = (\Phi\Phi^\dagger)^{-1}\ket{\phi_i}
\end{equation}
Here, $()^{-1}$ represents the Moore-Penrose inverse. And $\Phi$,$\tilde{\Phi}$ are matrices with $\ket{\phi_i}$,$\ket{\tilde{\phi_i}}$as columns respectively.
The Eq.~\ref{phi_i_tilde} is of great importance, as it is employed to obtain the reciprocal states of the states to be discriminated. These reciprocal states are then directly used in designing the POVM operators needed for the unambiguous discrimination of the $\ket{\phi_i}$ states.
The reciprocal states in this particular case are found as:
\begin{equation}
\begin{aligned}
    \ket{\tilde{\phi_1}} = \frac{1}{2a}\ket{0}^{\otimes n}\otimes\ket{0}
    +\frac{1}{2b}\ket{1}^{\otimes n}\otimes\ket{1}\\
    \ket{\tilde{\phi_2}} = \frac{1}{2a}\ket{0}^{\otimes n}\otimes\ket{0}
    -\frac{1}{2b}\ket{1}^{\otimes n}\otimes\ket{1}\\
    \ket{\tilde{\phi_3}} = \frac{1}{2a}\ket{1}^{\otimes n}\otimes\ket{0}
    +\frac{1}{2b}\ket{0}^{\otimes n}\otimes\ket{1}\\
    \ket{\tilde{\phi_3}} = \frac{1}{2a}\ket{1}^{\otimes n}\otimes\ket{0}
    -\frac{1}{2b}\ket{0}^{\otimes n}\otimes\ket{1}
\end{aligned}
\end{equation}
Let us consider the collection of m states, ($i=1,2,\dots, m$). Then (m+1) POVM elements are required. These $m$ elements correspond to conclusive detection of the $\ket{\phi_i}$ states and the remaining element corresponds to inconclusive result. The POVM operators are designed in the following manner for this specific case:
\begin{equation}\label{Pi i}
    \begin{aligned}
        \Pi_i = p_i\ket{\tilde{\phi_i}}\bra{\tilde{\phi_i}} 
    \end{aligned}
\end{equation}
\begin{equation}\label{Pi zero}
    \Pi_0 = I - \sum\Pi_i
\end{equation}
where $p_i$ represents the probabilities of detecting the states correctly. Considering the prior probabilities as $\eta_i$, the total conclusive probability is given by, 
\begin{equation}
    P_{con} = \sum \eta_i \bra{\phi_i}\Pi_i\ket{\phi_i} = \sum\eta_ip_i
\end{equation}
where we have employed the following relation:
\begin{equation}
    \langle{\phi_i}|{\tilde{\phi_k}}\rangle=\delta_{ik}
\end{equation}
In this case, the state $\ket{\phi_i}$ and prior probability $\eta_i$ satisfy the conditions for equal probability measurement \cite{chefles1998unambiguous} i.e. EPM. For optimal EPM, $p_i = p$ for $1\leq i\leq 4$. This in turn leads to p being the inverse of the maximum eigenvalue of the following frame operator
\begin{equation}
    S = \sum_{i=1}^4\ket{\tilde{\phi_i}}\bra{\tilde{\phi_i}}
\end{equation}
Thus we get $p = 2b^2$ and design the POVM operators as:
\begin{equation}\label{POVM}
    \begin{aligned}
        \Pi_i = 2b^2\ket{\tilde{\phi_i}}\bra{\tilde{\phi_i}} \\
        \Pi_0 = I - \sum\Pi_i
    \end{aligned}
\end{equation}
As per Eq. \ref{POVM}, The POVM operators are related to the coefficients of the quantum channel. The total probability of conclusive result is then given by 
\begin{equation}
    P_{con} = \sum\eta_ip_i = 4 \times \left( \frac{1}{4} \times 2b^2\right) = 2b^2
\end{equation}
Note that in case of maximally entangle quantum channel, $a= b=\frac{1}{\sqrt{2}}$. This implies that the state discrimination can be achieved with only PVM. Additionally the probability of conclusive result becomes $P_{con} = 2 \times \frac{1}{\sqrt{2}}$ = 1 which in turn leads to 100\% successful teleportation.

\subsubsection{Original state reconstruction}

The successful discrimination of states using POVM on Alice's qubits is crucial for the recovery of the original state. If the state is correctly detected Alice is required to relay 2 classical bits worth information to Bob through a classical channel. These cbits contain the result of measurement on nth and (n+1)th qubit. If the cbits are represented as $m_1m_2$, the correspondence relation between the POVM operators and the cbits may be formulated as,
\begin{equation}
    \Pi_1 \rightarrow 00\hspace{1cm}
    \Pi_2 \rightarrow 01\hspace{1cm}
    \Pi_3 \rightarrow 10\hspace{1cm}
    \Pi_4 \rightarrow 11\hspace{1cm}
\end{equation}
Depending on the cbits Alice relays to Bob, Bob is now required to do appropriate local operation on his qubits in order to reconstruct the original state. The following table shows the unitary operations  required for each of the four possible cases:
\begin{table}[h]
    \centering
    \begin{tabular}{|c|c|c|}
        \hline
        $m_1m_2$ & State on Bob's qubits & Local operation \\ \hline
        00 & $(\alpha\ket{0}^{\otimes n} + \beta\ket{1}^{\otimes n})_{n+2,n+3,\dots,2n+1}$ & $I_{n+2} \otimes \dots \otimes I_{2n+1}$ \\ \hline
        01 & $(\alpha\ket{0}^{\otimes n} - \beta\ket{1}^{\otimes n})_{n+2,n+3,\dots,2n+1}$ & $I_{n+2} \otimes \dots \otimes I_{2n} \otimes Z_{2n+1}$ \\ \hline
        10 & $(\beta\ket{0}^{\otimes n} + \alpha\ket{1}^{\otimes n})_{n+2,n+3,\dots,2n+1}$ & $X_{n+2} \otimes \dots \otimes X_{2n+1}$ \\ \hline
        11 & $(\beta\ket{0}^{\otimes n} - \alpha\ket{1}^{\otimes n})_{n+2,n+3,\dots,2n+1}$ & $X_{n+2} \otimes \dots \otimes (X Z)_{2n+1}$ \\ \hline
    \end{tabular}
    \caption{List of appropriate unitary operations}
    \label{tab:unitary}
\end{table} 

\noindent The unitary operations needed to recover the original state $\ket{\chi_0}$ in all four possible cases, as shown in Table~\ref{tab:unitary}, can be written as:

\begin{equation}\label{gen unitary}
    T = X_{n+2}^{m_1}\otimes X_{n+3}^{m_1}\otimes \dots \otimes (X^{m_1} Z^{m_2})_{2n+1}
\end{equation}
where X = $\big(\begin{smallmatrix}
  0 & 1\\
  1 & 0
\end{smallmatrix}\big)$  and Z = $\big(\begin{smallmatrix}
  1 & 0\\
  0 & -1
\end{smallmatrix}\big)$ are Pauli matrices.

\noindent The Eq.~\ref{gen unitary} shows that the only requirement for finding which local operation Bob must apply is knowing the two cbits $m_1$ and $m_2$. Thus applying the unitary operation $T$ on Bob's qubits ensures the successful recovery of the original state and hence concludes the teleportation scheme successfully.

\subsection{Generalization for Broader Classes of Entangled States}

\begin{figure}[h]
    \centering
    \includegraphics[width=01\linewidth]{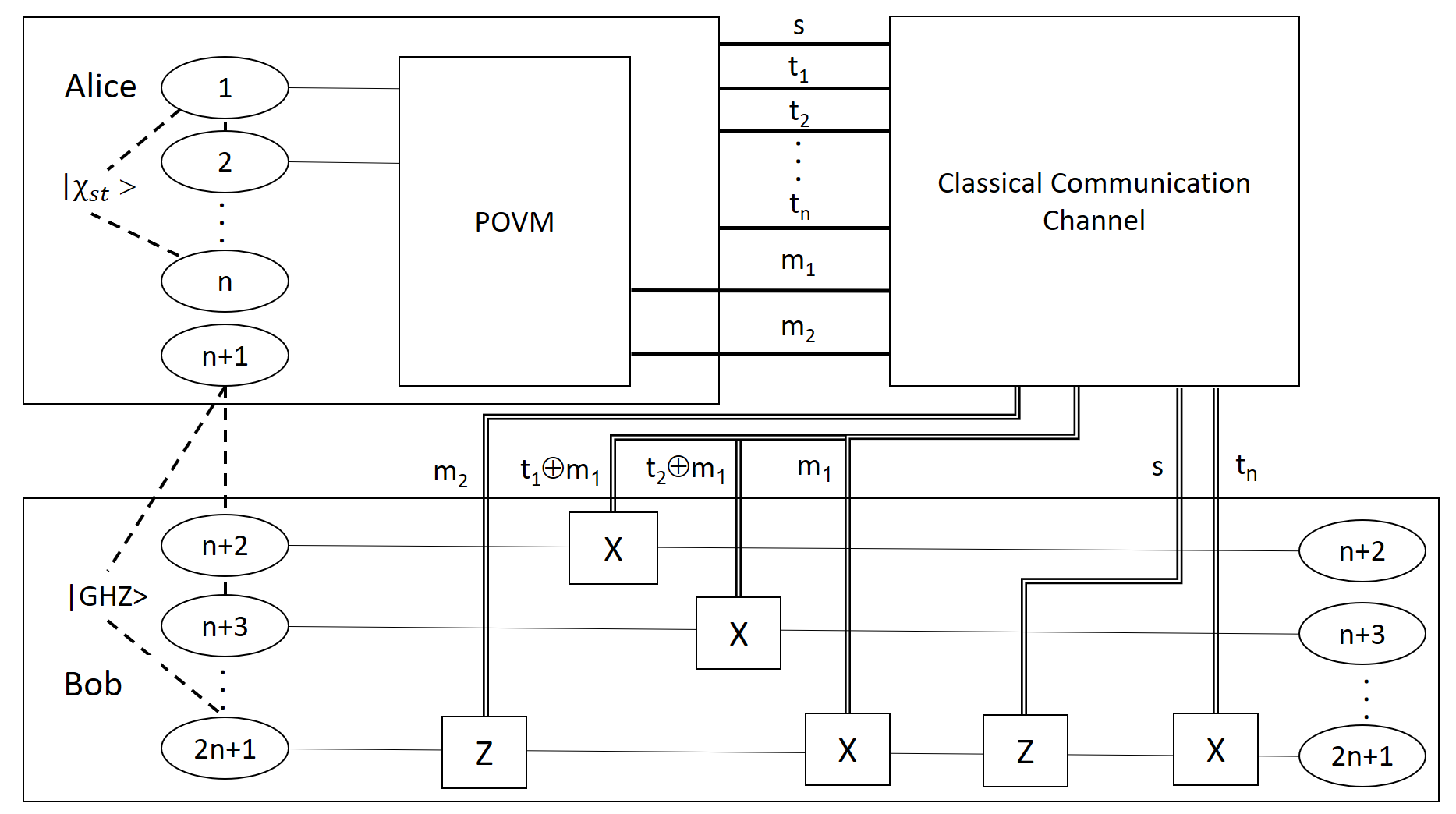}
    \caption{Quantum teleportation of a more generalized n-qubit entangled state. Alice holds an unknown n-qubit state $\ket{\chi_{st}} = \alpha \ket{t_1t_2...t_n} + (-1)^{s} \beta \ket{t_1^\prime t_2^\prime...t_n^\prime}$ and one qubit from a shared partially entangled GHZ channel. She performs a POVM on her (n+1) qubits and sends the classical outcomes $m_1$, $m_2$, and additional bits $t_1, t_2,\dots,t_n, s$ to Bob via a classical communication channel. Bob, holding the remaining n qubits, applies X and Z gates based on the received bits as per Eq.~\ref{t_prime} to reconstruct $\ket{\chi_{st}}$.}
    \label{fig_chi_st}
\end{figure}

We consider the generalized input state is given by,
\begin{equation}\label{chi_st}
    \ket{\chi_{st}} = \alpha \ket{t_1t_2...t_n} + (-1)^{s} \beta \ket{t_1^\prime t_2^\prime...t_n^\prime}
\end{equation}
Here, $t_i,s$ = 0 or 1 and $t^\prime_i = 1 - t_i$. 
One can easily deduce that the input state, although generalized in form, can be taken as a single logical qubit. However, we choose to treat this initial state as a whole for the following formulation in order to align with our motivation i.e. resilience against bit-flip type error. To intuitively realize the protocol, one might follow the Figure~\ref{fig_chi_st}.
For such a generalization, we highly rely on the following relation,
\begin{equation}\label{chi_st_relation}
    \ket{\chi_{st}} = U_{st}\ket{\chi_0}
\end{equation}
The unitary $U_{st}$ can be decomposed in terms of only local operations i.e. $Z$ and $X$ gate. 
\begin{equation}\label{U st}
    U_{st} = X_{(n+2)}^{t_1}\otimes X_{(n+3)}^{t_2} \otimes \dots \otimes X_{(2n+1)}^{t_n} Z_{(2n+1)}^{s} 
\end{equation}
 Again, we consider pure partially entangled GHZ state as the quantum channel. Now the combined state of the system can be written as,
\begin{equation}
\begin{aligned}
    \ket{\psi_{sys}^\prime} &= \ket{\chi_{st}}_{1,2,...,n}\otimes\ket{GHZ}_{n+1,n+2,...,2n+1} \\
    &= U_{st}\ket{\chi_{0}}_{1,2,...,n}\otimes\ket{GHZ}_{n+1,n+2,...,2n+1}
\end{aligned}
\end{equation}
Expanding and rearranging the terms,
\begin{equation}
    \begin{aligned}
        \ket{\psi_{sys}^{\prime}} &= U_{st}(\alpha\ket{0}^{\otimes n} + \beta\ket{1}^{\otimes n})_{1,2,\dots,n} 
        \otimes (a\ket{0}^{\otimes n+1} + b\ket{1}^{\otimes n+1})_{n+1,n+2,\dots,2n+1}\\ 
        &= \frac{1}{2} (U_{st} \otimes I) (\ket{\phi_1})_{1,2,\dots,n+1} 
        \otimes (\alpha\ket{0}^{\otimes n} + \beta\ket{1}^{\otimes n})_{n+2,n+3,\dots,2n+1} \\
        &\quad + \frac{1}{2} (U_{st} \otimes I) (\ket{\phi_2})_{1,2,\dots,n+1} 
        \otimes (\alpha\ket{0}^{\otimes n} - \beta\ket{1}^{\otimes n})_{n+2,n+3,\dots,2n+1} \\
        &\quad + \frac{1}{2} (U_{st} \otimes I) (\ket{\phi_3})_{1,2,\dots,n+1} 
        \otimes (\beta\ket{0}^{\otimes n} + \alpha\ket{1}^{\otimes n})_{n+2,n+3,\dots,2n+1} \\
        &\quad + \frac{1}{2} (U_{st} \otimes I) (\ket{\phi_4})_{1,2,\dots,n+1} 
        \otimes (\beta\ket{0}^{\otimes n} - \alpha\ket{1}^{\otimes n})_{n+2,n+3,\dots,2n+1}.
    \end{aligned}
\end{equation}

Let us define a new unitary $U^{\prime}$ as:
\begin{equation}\label{U_prime}
    U^{\prime} = U_{st}\otimes I
\end{equation}
The new set of states $\ket{\phi_i^\prime}$ and the previously discussed states $\ket{\phi_i}$ have the following relation,
\begin{equation}\label{phi prime}
    \ket{\phi_i^\prime} = U^\prime \ket{\phi_i}
\end{equation}
 By taking \( \ket{\phi_i^\prime} \) as columns, the matrix ${\Phi}^\prime$ can be constructed. We additionally know that $\Phi^\prime = U^\prime \Phi$. The matrix \( \tilde{\Phi}^\prime \) can be obtained easily through Eq.~\ref{phi_i_tilde} so that,

\begin{equation}
    \begin{aligned}
        \ket{\tilde{\phi}^\prime} 
        &= (\Phi^{\prime} \Phi^{\prime\dagger})^{-1} \ket{\phi^\prime} \\
        &= \big((U^\prime \Phi) (U^\prime \Phi)^{\dagger}\big)^{-1} U^\prime \ket{\phi} \\
        &= ( U^{\prime} \Phi  \Phi^{\dagger} U^{\prime\dagger})^{-1} U^\prime \ket{\phi} \\ 
        &= (U^{\prime\dagger})^{-1}((\Phi \Phi^\dagger )^{-1}(U^{\prime})^{-1} U^\prime \ket{\phi}\\
        &= U^{\prime}(\Phi \Phi^{\dagger})^{-1}\ket{\phi}
    \end{aligned}
\end{equation}
This implies that the rigorous calculation of finding the reciprocal states for the set of state $\ket{\phi_i^\prime}$ can be bypassed and instead be written as:
\begin{equation}\label{phi prime tilde}
    \ket{\tilde{\phi_i^\prime}} = U^\prime \ket{\tilde{\phi_i}}
\end{equation}
Since the states $\ket{\tilde{\phi_i}}$ are known for the case of $\ket{\chi_0}$, this saves us a lot of steps in designing the optimal POVM for state discrimination.

After optimal state discrimination, the state $\ket{\chi_{st}}$ can be reconstructed by finding out the appropriate unitary operations needed to be applied on Bob's qubits. However, we use a different approach. We stack the unitary operations over the operations done in the previous protocol for teleportation of $\ket{\chi_{0}}$ state as shown in Eq.~\ref{gen unitary}. Thus the unitary operation needed for the reconstruction of the $\ket{\chi_{st}}$ can be written in the following generalized form:
\begin{equation}\label{t_prime}
\begin{aligned}
     T^\prime &= U_{st}T \\
    &= (X_{n+2}^{t_1}\otimes X_{n+3}^{t_2} \otimes \dots \otimes (X^{t_n} Z^{s})_{2n+1}) \\
    &\quad(X_{n+2}^{m_1}\otimes X_{n+3}^{m_1} \dots \otimes (X^{m_1} Z^{m_2})_{2n+1})\\
     &= X_{n+2}^{t_{1} \oplus m_1 } \otimes X_{n+3}^{t_{2} \oplus m_1}\otimes \dots \otimes (X^{t_{n}} Z^{s} X^{m_1} Z^{m_2})_{2n+1}
\end{aligned}
\end{equation}

\begin{table}[!htb]
    \centering
    \begin{tabular}{|c|c|c|}
        \hline
        $m_1m_2$ & State on Bob's qubits & Local operation \\ \hline
        00 & $(\alpha\ket{0}^{\otimes n} + \beta\ket{1}^{\otimes n})$ & $X_{n+2}^{t_{1}} \otimes X_{n+3}^{t_{2}}\otimes \dots \otimes (X^{t_{n}} Z^{s} )_{2n+1}$ \\ \hline
        01 & $(\alpha\ket{0}^{\otimes n} - \beta\ket{1}^{\otimes n})$ & $X_{n+2}^{t_{1}} \otimes X_{n+3}^{t_{2}}\otimes \dots \otimes (X^{t_{n}} Z^{s} Z)_{2n+1}$ \\ \hline
        10 & $(\beta\ket{0}^{\otimes n} + \alpha\ket{1}^{\otimes n})$ & $X_{n+2}^{t_{1}^{\prime}} \otimes X_{n+3}^{t_{2}^{\prime}}\otimes \dots \otimes (X^{t_{n}} Z^{s} X )_{2n+1}$ \\ \hline
        11 & $(\beta\ket{0}^{\otimes n} - \alpha\ket{1}^{\otimes n})$ & $X_{n+2}^{t_{1}^{\prime}} \otimes X_{n+3}^{t_{2}^{\prime}}\otimes \dots \otimes (X^{t_{n}} Z^{s} X Z)_{2n+1}$ \\ \hline
    \end{tabular}
    \caption{List of appropriate unitary operations}
    \label{tab: psi_st unitary}
\end{table} 

Note that by only knowing the $m_1$, $m_2$, $s$ and $t_i$, one can easily find out the specific expression of $T^{\prime}$ as shown in the Table~\ref{tab: psi_st unitary}. In this case, Alice is obligated to relay $(n+3)$ bit worth classical information to Bob. Thus, applying this unitary $T^{\prime}$ on Bob's end successfully completes the scheme. The increase in number of cbits enables this protocol to accommodate $2^{n+1}$ type of entangled states.

\subsection{Algorithmic Implementation}
The following algorithm implements a protocol for teleporting a generalized $n$-qubit entangled state, $\ket{\chi_{st}} = \alpha \ket{t_1 t_2 \dots t_n} + (-1)^s \beta \ket{t_1' t_2' \dots t_n'}$, using a partially entangled GHZ state as the quantum channel. This approach leverages an optimal Positive Operator-Valued Measure (POVM) to address the challenge of distinguishing non-orthogonal states, incorporating eigenvalue decomposition for inconclusive outcomes and Naimark's dilation theorem for efficient simulation. The result is a resource-efficient method for quantum state teleportation, detailed in the Algorithm 1.

\begin{algorithm*}[!t]
\caption{Teleportation of Generalized $n$-Qubit Entangled State}
\begin{algorithmic}[1]
\State \textbf{Input:} $n$, $s$, $t = [t_1, t_2, \dots, t_n]$, $a$, $b$
\State \textbf{Output:} Teleported state on Bob's qubits

\State Initialize quantum circuit with $2n+1$ qubits and $n+1$ classical bits

\State Prepare input state $\ket{\chi_{st}} = \alpha \ket{t_1 t_2 \dots t_n} + (-1)^s \beta \ket{t_1' t_2' \dots t_n'}$ on qubits 1 to $n$

\State Construct unitary $U_{st} = X^{t_1}_{n+2} \otimes X^{t_2}_{n+3} \otimes \cdots \otimes X^{t_n}_{2n} \otimes Z^s_{2n+1}$

\State Prepare partially entangled GHZ state $\ket{\text{GHZ}} = a\ket{0}^{\otimes (n+1)} + b\ket{1}^{\otimes (n+1)}$ on qubits $n+1$ to $2n+1$

\State Form total state $\ket{\psi'_{\text{sys}}} = \ket{\chi_{st}}_{1,\dots,n} \otimes \ket{\text{GHZ}}_{n+1,\dots,2n+1}$

\State Define $U' = U_{st} \otimes I$

\State Compute reciprocal states $\ket{\tilde{\phi_i'}} = U' \ket{\tilde{\phi_i}}$ for $i=1,2,3,4$

\State Construct POVM elements:
\State \quad $\Pi_i = 2b^2 \ket{\tilde{\phi_i'}}\bra{\tilde{\phi_i'}}$ for $i=1,2,3,4$
\State \quad $\Pi_0 = I - \sum_{i=1}^4 \Pi_i$

\State Perform state discrimination on qubits 1 to $n+1$ using POVM $\{\Pi_0, \Pi_1, \Pi_2, \Pi_3, \Pi_4\}$

\If {measurement outcome is $\Pi_0$}
    \State Restart the protocol
\Else
    \State Measurement results on qubits $n$ and $n+1$ provide $m_1, m_2$
    \State Send classical bits $m_1, m_2$ to Bob
    \State Bob applies unitary $T' = X^{t_1 \oplus m_1}_{n+2} \otimes X^{t_2 \oplus m_1}_{n+3} \otimes \cdots \otimes (X^{t_n} Z^s X^{m_1} Z^{m_2})_{2n+1}$ on his qubits
\EndIf

\State \textbf{Return:} Bob's qubits $n+2$ to $2n+1$ holding the teleported state
\end{algorithmic}
\end{algorithm*}

\subsection{Illustrative Example: Teleportation of a 3-Qubit State}

Let us consider the state to be teleported as:
\begin{equation}
    \ket{\chi_{st}^\prime} = \alpha \ket{101} - \beta \ket{010}
\end{equation}
Here, $n$=3, $s$=1, $t_1$=1, $t_2$=0, $t_3$=1 and accordingly $t_1^{\prime}$=0, $t_2^{\prime}$=1, $t_3^{\prime}$=0
Thus the expression of $U_{st}$ in this case is given by,
\begin{equation}
\begin{aligned}
    U_{st}&=X_{5}^{t_1}\otimes X_{6}^{t_2}\otimes (X^{t_3} Z^{s})_7  \\
    &=X_{5}^{1} \otimes X_{6}^{0}\otimes (X^{1} Z^{1})_7\\
    &=X_{5} \otimes I_{6}\otimes (X Z)_7
\end{aligned}
\end{equation}
This in turn leads to,
\begin{equation}
\begin{aligned}
    U^{\prime} &= U_{st} \otimes I\\
    &= X_{5} \otimes I_{6}\otimes (X Z)_7 \otimes I
\end{aligned}
\end{equation}
By making use of Eq.~\ref{phi prime} and Eq.~\ref{phi prime tilde} we find the reciprocal states $\ket{\tilde{\phi^\prime_i}}$ where $i$=1,2,3,4.
Now using these states, optimal POVM can be designed for unambiguous state discrimination. 

\begin{figure}[h]
    \centering
    \includegraphics[width=01\linewidth]{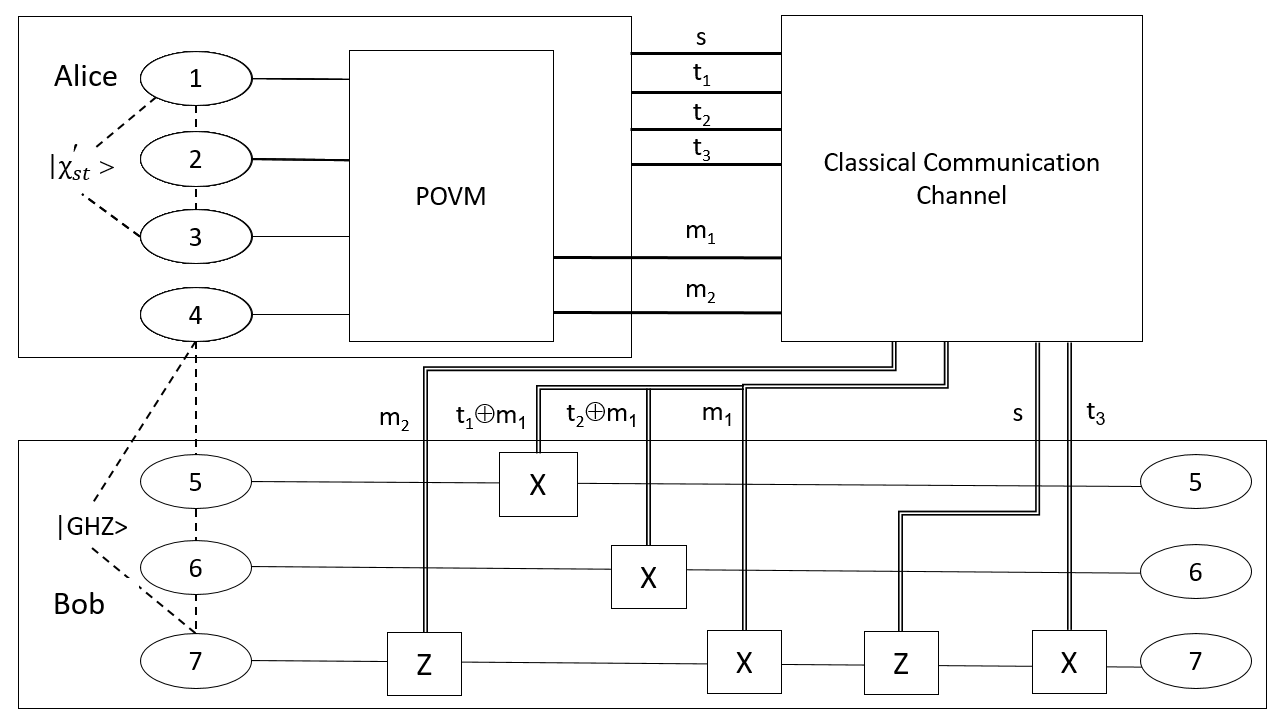}
    \caption{Quantum teleportation of a 3-qubit entangled state. Alice holds an unknown 3-qubit state $\ket{\chi_{st}^\prime} = \alpha \ket{101} - \beta \ket{010}$ and one qubit from a shared 4-qubit partially entangled GHZ state. After performing a POVM on these qubits, she sends the classical bits $m_1,m_2,t_1=1,t_2=0,t_3=1$ and $s=1$ to Bob. Bob, holding the remaining 3 qubits, applies X and Z gates based on the received bits as per Eq.~\ref{t_prime_expression} to reconstruct $\ket{\chi_{st}^{\prime}}$.}
    \label{fig_3_qubit}
\end{figure}

The entire steps to implement the protocol for the 3 qubit input state within the specific scope of the proposed scheme can be seen in the Figure~\ref{fig_3_qubit}.

If Alice sends the relevant classical bits i.e. measurement results on qubits 3,4, then Bob just needs to apply the unitary operation $T^\prime$ on his qubits. In this case the expression for $T^\prime$ may be found as per Eq.~\ref{t_prime},
\begin{equation}\label{t_prime_expression}
\begin{aligned}
    T^\prime
    &= X_{5}^{t_1 \oplus m_1 } \otimes X_{6}^{t_2 \oplus m_1}\otimes (X^{t_3} Z^{s} X^{m_1} Z^{m_2})_7\\
    &= X_{5}^{1 \oplus m_1 } \otimes X_{6}^{0 \oplus m_1}\otimes (X^{1} Z^{1} X^{m_1} Z^{m_2})_7\\
    &= X_{5}^{1 \oplus m_1 } \otimes X_{6}^{m_1}\otimes (X Z X^{m_1} Z^{m_2})_7
\end{aligned}
\end{equation}
The appropriate unitary operations in each of the four possible cases may be listed as:
\begin{table}[H]
    \centering
    \begin{tabular}{|c|c|c|}
        \hline
        $m_1m_2$ & State on Bob's qubits & Local operation \\ \hline
        00 & $(\alpha\ket{000} + \beta\ket{111})_{5,6,7}$ & $X_{5} \otimes I_{6}\otimes (X Z)_7$ \\ \hline
        01 & $(\alpha\ket{000} - \beta\ket{111})_{5,6,7}$ & $X_{5} \otimes I_{6}\otimes X_{7} $ \\ \hline 
        10 & $(\beta\ket{000} + \alpha\ket{111})_{5,6,7}$ & $I_{5} \otimes X_{6}\otimes (X Z X)_7$ \\ \hline 
        11 & $(\beta\ket{000} - \alpha\ket{111})_{5,6,7}$ & $I_{5} \otimes X_{6}\otimes \left( X Z X Z \right)_7$ \\ \hline
    \end{tabular}
    \caption{List of appropriate unitary operations}
    \label{tab:3 qubit unitary}
\end{table}

\subsection{Efficient POVM Implementation and Cost Analysis}

A crucial aspect of the protocol's viability is the implementation cost of the joint measurement on Alice's $n+1$ qubits. A naive implementation on an arbitrary $(n+1)$-qubit state would require resources that scale exponentially with $n$, rendering the protocol impractical. However, the specific structure of the logical qubit basis states, $\ket{0_L}$ and $\ket{1_L}$, allows for a highly efficient measurement process.

The key to this efficiency is that the measurement only needs to distinguish between the two orthogonal logical basis states. This does not require a complex, simultaneous measurement on all qubits. Instead, Alice can first perform a simple preparatory circuit of $n-1$ CNOT gates. The effect of this circuit is to consolidate the information that distinguishes $\ket{0_L}$ from $\ket{1_L}$ onto a single, representative qubit, while the other $n-1$ qubits are reset to the known $\ket{0}$ state. Consequently, the complex part of the POVM measurement can now ignore these other qubits and only needs to act on the two that matter: the single representative qubit and the channel qubit. Therefore, the total gate cost for Alice's measurement scales \textbf{linearly with n}, for any reasonably large $n$.

\section{Protocol Analysis, Scope, and Applications}

This section analyzes the performance of the proposed protocol within its intended application of fault-tolerant state transfer. We begin by examining the protocol's probabilistic nature, which is tied to the quality of the quantum channel. We then present a detailed fidelity comparison against the conventional alternative, which serves as the core justification for our approach.

\subsection{Success Probability and Channel Quality}

\begin{figure}
    \centering
    \includegraphics[width=0.75\linewidth]{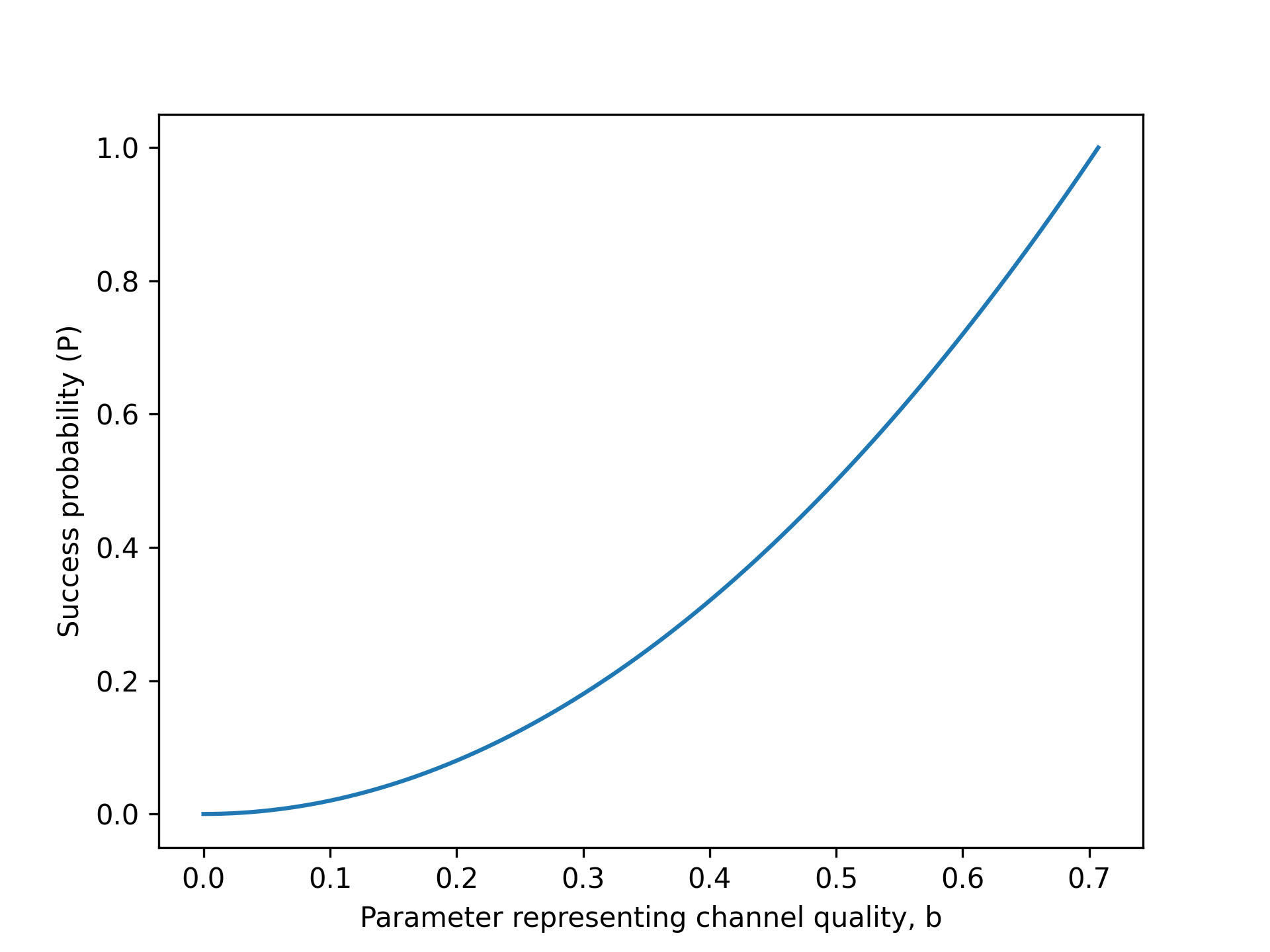}
    \caption{Graph showing the variation of success probability depending on the channel quality represented by the parameter b. According to the rule a>b for this specific protocol, the value of b ranges from 0 to $1/\sqrt{2}$. And the probability of successful teleportation $P=2b^2$ inherently implying that for maximally entangled channel i.e. when b=$1/\sqrt{2}$ the success probability i.e. fidelity becomes 1.}
    \label{fig:P_vs_b}
\end{figure}

 The use of an optimal POVM on a partially entangled channel means that the teleportation is probabilistic. The protocol succeeds only when the POVM yields a conclusive measurement outcome; an inconclusive result requires the protocol to be restarted. As derived in Eq. (3.15), the total probability of a conclusive result is given by $P_{\text{con}} = 2b^2$
where $|b|$ is the coefficient of the $\ket{1}^{\otimes n+1}$ term in the shared GHZ channel. This success probability is therefore a direct and measurable indicator of the quality of the entanglement resource as shown in Figure~\ref{fig:P_vs_b}. 
For a maximally entangled channel where $|a|=|b|=1/\sqrt{2}$, the success probability becomes $P_{\text{con}} = 1$, and the teleportation is deterministic. This inherent trade-off between channel quality and success rate is a key feature of operating in realistic, noisy quantum networks.

\subsection{Comparative Fidelity Analysis}

The core advantage of our protocol lies in avoiding the gate errors associated with a conventional decode-teleport-re-encode sequence. We explicitly compare the two pathways:

\begin{description}

    \item[Direct Logical Teleportation (This Work)] The logical qubit is teleported in its encoded form. The protocol's fidelity is limited by channel noise and POVM imperfections, but it inherits the resilience of the error-correcting code.

    \item[Decode-Teleport-Re-encode (Conventional Alternative)] This pathway introduces three sequential error sources: (1) gate errors during Alice's decoding circuit, (2) infidelity in the bare-qubit teleportation, and (3) gate errors during Bob's re-encoding circuit.

\end{description}

A single error during Alice's decoding step is catastrophic, as it corrupts the quantum \mbox{information} \textit{before} it is protected.

If we denote the two-qubit gate error rate as $p_g$, the probability of such a decoding failure is approximately $(n-1)p_g$. To quantify this trade-off, we consider a scenario with $n=5$ qubits. For Pathway B, using a typical two-qubit gate error rate of $p_g = 0.5\%$ for 2025-era superconducting platforms, the infidelity has a hard floor of:

$$P_{\text{decode}} \approx (n-1) \times p_g = (5-1) \times 0.005 = 0.02 = \mathbf{2\%}$$

This 2\% failure probability represents a significant initial hurdle before teleportation even occurs.

In contrast, our direct protocol (Pathway A) inherits the error-suppressing capabilities of the $[[5,1,1]]$ repetition code. Assuming a physical bit-flip rate of $p=1\%$ on each qubit rail, the logical error rate $P_L$ is the sum of probabilities for 3, 4, or 5 physical errors occurring:
$$P_L = \sum_{k=3}^{5} \binom{5}{k} p^k (1-p)^{5-k} \approx 0.000985\%$$

This logical error rate is more than three orders of magnitude lower than the failure floor of the conventional alternative, demonstrating a clear and significant fidelity advantage for our direct teleportation approach.

\subsection{Integration of the proposed scheme in quantum hop-by-hop communication}

In the proposed teleportation scheme, only Alice i.e. the sender requires complete knowledge of the quantum channel. Bob i.e. the receiver is not obligated to be aware of the coefficients of the quantum channel. Note that our proposed scheme is applicable in a limited scope. However, in case of teleporting any entangled state within the scope of the proposed scheme, the sender can choose which of the two protocols(our proposed scheme and the typical scheme using Bell basis measurement) they wish to follow. Both the scheme can be utilized alternately without disrupting the quantum state to be teleported. We take this implication a bit further and find an opportunity to increase security in quantum hop-by-hop communication.

\begin{figure}[h]
    \centering
    \includegraphics[width=01\linewidth]{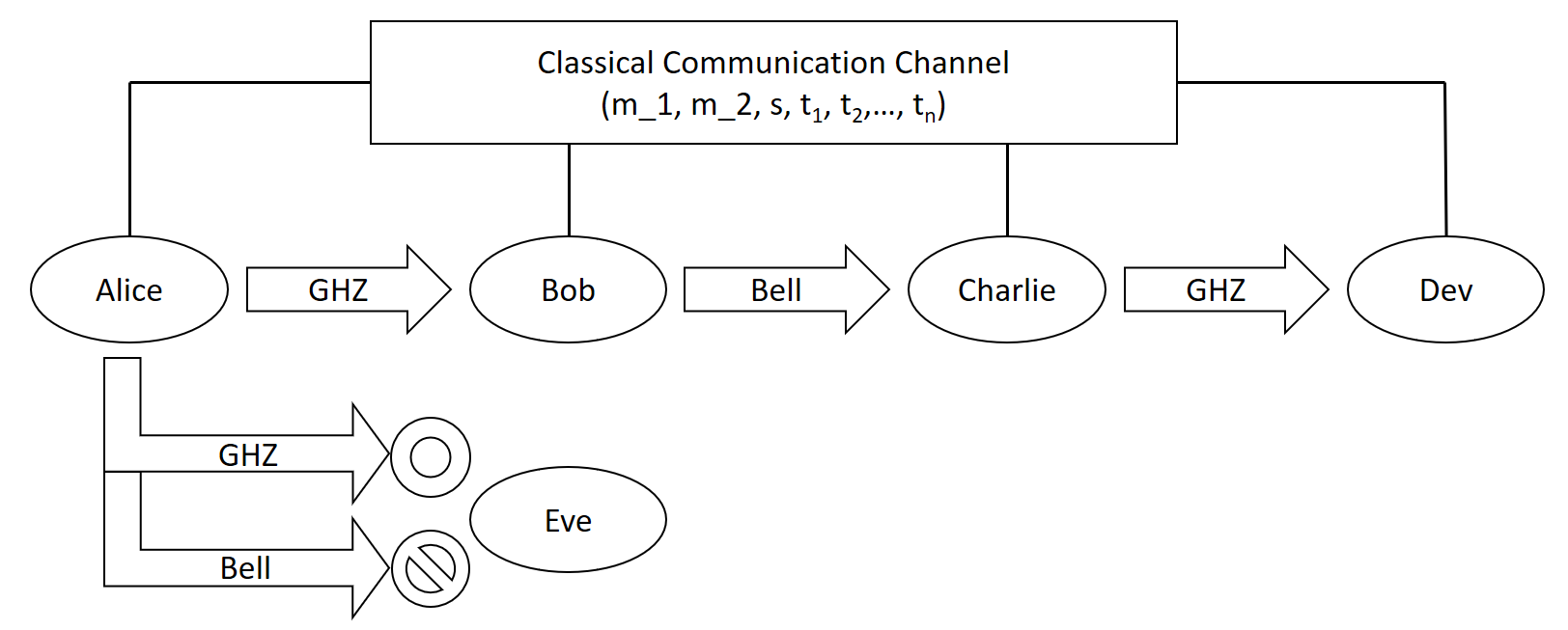}
    \caption{Quantum hop-by-hop communication with enhanced security; Nodes communicate using either GHZ or Bell state entanglement, chosen randomly, when teleporting entangled states within the scope of the proposed scheme. Classical information ($m_1,m_2,s,t_1,t_2,\dots,t_n$) is shared over a public channel but does not reveal any details about the type of quantum channel used between any two immediate nodes. An eavesdropper (Eve) therefore has only a 50\% chance of correctly identifying the entanglement type (GHZ or Bell) shared between any two nodes, such as Alice and Bob, enhancing the protocol’s security compared to using only Bell-type correlations.}
    \label{fig:integration}
\end{figure}

\par Let us consider four nodes named Alice, Bob, Charlie and Dev. Alice wishes to send a quantum state of the form $\alpha \ket{t_1t_2\dots t_n}+(-1)^s \beta \ket{t_1^{\prime}t_2^{\prime}\dots t_n^{\prime}}$ to Dev through the intermediate nodes Bob and Charlie. Now all of the nodes have the choice of following either the proposed scheme or the typical scheme that utilizes Bell basis measurement. If an eavesdropper Eve wishes to intercept the quantum information, she is at first obligated to know all the information relayed through the classical channel. These information include $m_1,m_2,s$ and $t_i$s. After obtaining that information, Eve would need to entangle her own qubits with the nth qubit of Alice or any sender node in the communication channel. However, the information relayed through classical channel does not contain any 
meaningful information about the quantum channel used between any of the two immediate nodes. This implies that Eve has a 50 percent chance of correctly guessing the type of quantum channel used between any two immediate nodes. On the other hand, if the nodes in the hop-by-hop communication system rely on only one type of quantum channel then Eve's chances of correctly guessing becomes significantly higher. This shows that by alternating between the proposed and typical teleportation protocol randomly, we can enhance the security of such quantum hop-by-hop communication.
\newline
\newline

\section{Discussion and conclusions}

\par This proposed scheme is applicable in teleporting n qubit quantum states via partially entangled quantum channel like in \cite{wang2018two} only in cases where the input quantum states can be considered as equivalent to a logical qubit. This leads to the usage of (n+1) qubit quantum channel. The protocol utilizes a partially entangled GHZ state as quantum channel for introducing experimental error and decoherence factors. Additionally, we choose to formulate the reciprocal states following Yonina C Eldar \cite{eldar2003semidefinite}  instead of employing analogous reciprocal states used in \cite{wang2018two}. The optimal POVM designed for unambiguous state discrimination leads to inconclusive results in some cases. In such cases, Alice would start the protocol anew. In case of teleporting a state as shown in Eq.~\ref{chi_st}, one can use the unitary as shown in Eq.~\ref{U st}. This is limited to the case when the unitary needed to construct the initial state as per Eq.~\ref{chi_st_relation} is composed of only Z and X quantum gates. In such a case, the receiver would only need to apply the unitary operation denoted by $T^{\prime}$ as shown in Eq.~\ref{t_prime}. The success probability of this teleportation scheme depends on the quality of the channel used, more specifically, on the parameter b of the partially entangled GHZ channel as shown in Figure~\ref{fig:P_vs_b}.
\begin{figure}
    \centering
    \includegraphics[width=0.75\linewidth]{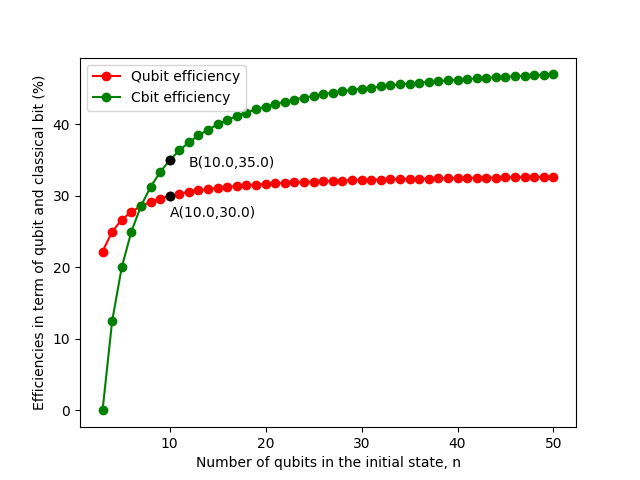}
    \caption{Graph showing the rising qubit and cbit efficiency with increasing number of qubits in the input entangled state where a single logical qubit is encoded. As n becomes sufficiently large, the qubit efficiency tends to 33.33\% and the cbit efficiency reaches almost 50\%. The points A(10,30) and B(10,35) show that the qubit and cbit efficiencies for $n=10$ are 30\%,35\%.}
    \label{fig:efficiency_vs_n}
\end{figure}
\par
The proposed teleportation scheme is capable of teleporting specific type of $n$-qubit entangled states more efficiently than the typical teleportation scheme utilizing Bell Basis Measurement. In the proposed protocol, for $n$-qubit state teleportation, one would require $2n+1$ qubits and $n+3$ classical bits in total. Conversely in the typical protocol using Bell basis, one would need $3n$ qubits and $2n$ classical bits for teleportation of $n$-qubit state. For the purpose of comparison we define the following parameters representing the efficiencies in terms of qubits and classical bits:
$$
\text{Relative efficiency (qubits), } \eta_{q} = \frac{3n - (2n + 1)}{3n} \times 100\% =\left(\frac{1}{3} - \frac{1}{3n} \right) \times 100\%
$$
$$
\text{Relative efficiency (cbits), } \eta_{c} = \frac{2n - (n + 3)}{2n} \times 100\%= \left(\frac{1}{2} - \frac{3}{2n} \right) \times 100\%
$$
This implies that our proposed protocol is $(\frac{1}{3}-\frac{1}{3n})\times100\%$ more efficient in terms of qubits and $(\frac{1}{2}-\frac{3}{2n})\times100\%$ more efficient in terms of classical bits than the typical protocol. As $n\to \infty$, the relative qubit and cbit efficiency approaches 33.33\% and 50\% respectively. However, even in small value range of n, the efficiency is quite high. For example, at $n=10$, qubit and cbit efficiencies are 30\% and 35\% as seen from the points A and B in Figure~\ref{fig:efficiency_vs_n}. The proposed protocol is not applicable for the teleportation of any general $n$-qubit entangled state. But, it proves to be more suitable for teleportation of a type of entangled quantum states of the form $\alpha \ket{t_1t_2\dots t_n}+(-1)^s \beta \ket{t_1^{\prime}t_2^{\prime}\dots t_n^{\prime}}$ which is equivalent to a logical qubit pre-encoded in n physical qubits. In the future, the exact noise resilience of this work might be investigated and further improved upon to incorporate two qubit gate error correction. The security of the classical messages sent has not been the focus of this work. However, significant improvements might be achieved in this regard employing certain strategies e.g. zero knowledge proof, quantum secret sharing of classical messages etc.
\par
The main advantage of this proposed scheme is its error suppressing capabilities when teleporting a logical qubit encoded in n physical qubits. As shown in sub-section 4.2, the logical error rate in our scheme is more than three orders of magnitude lower than the failure floor of the popular alternative. This illustrates a significant advantage in case of fidelity of our direct teleportation approach. Another advantage of this scheme is that the receiver is unaware of the coefficients of the quantum channel. This implies that the scheme is applicable interchangeably with the typical teleportation scheme using Bell basis measurement. Since the receiver does not require any knowledge of the quantum channel's coefficients, the sender can choose whether to follow the proposed scheme or the typical scheme. This shows a complete formulation of the quantum chain communication between any number of nodes.


\section{Data Availability}
The data that support the findings of this study are available from the corresponding author
upon reasonable request.

\section{Author Declarations }
\subsection{Conflict of Interest}
The authors have no conflicts to disclose.

\printbibliography
\end{document}